\documentclass[final,preprint,aps,showpacs,tightenlines]{revtex4}
\usepackage{graphicx}
\usepackage{bm}
\usepackage{dcolumn}

\begin{document}

\bibliographystyle{prsty}

%%%%%%%%%%%%%%%%%%%%%%%%%%%%%%%%%%%%%%%%%%%%%%%%%%%%%%%%%%%%%%%%%%%%%%%%%%%

\preprint{\vbox{\hbox{PNU-NTG-03/2000} \hbox{RUB-TPII-16/00} }}
\title{Strange form factors \\ in the context of SAMPLE, HAPPEX, \\ and A4 experiments}

\author{Ant\'onio Silva$^{(1,2)}$
\footnote{E-mail address: Antonio.Silva@tp2.ruhr-uni-bochum.de},
Hyun-Chul Kim$^{(3)}$
\footnote{E-mail address: hchkim@pnu.edu},
and Klaus Goeke$^{(1)}$
\footnote{E-mail address: Klaus.Goeke@tp2.ruhr-uni-bochum.de}}

\affiliation{(1) Institut f\"ur Theoretische  Physik  II, \\  
Ruhr-Universit\" at Bochum, \\
 D--44780 Bochum, Germany  \\
(2) Departamento de F\'\i sica and Centro de F\'\i sica Computacional,\\
Universidade de Coimbra,\\
P-3000 Coimbra, Portugal \\
(3) Department of Physics,
Pusan National University,\\
609-735 Pusan, Republic of Korea\\
       }
\date{September 2001}

\begin{abstract}
The strange properties of the nucleon are investigated within the framework of 
the SU(3) chiral quark-soliton model assuming isospin symmetry and applying
the symmetry-conserving SU(3) quantization.
We present the form factors $G^0_{E,M}(Q^2)$, $G^Z_M(Q^2)$ and the electric and 
magnetic strange form factors $G^{\rm s}_{E,M}(Q^2)$ incorporating pion and kaon asymptotics.  
The results show a fairly good agreement with the recent experimental data 
from the SAMPLE and HAPPEX collaborations.
We also present predictions for future measurements including the A4 experiment
at MAMI (Mainz).
\end{abstract}

\maketitle

%111111111111111111111111111111111111111111111111111111111111111111

{\bf 1.}  The understanding of the role that the sea quarks play 
          in the properties of the nucleon and other hadrons is of fundamental importance 
to the description of their internal dynamics. 
With this respect, the determination
of the strange contribution to the properties of the nucleon, for instance to its
magnetic moment, charge and magnetic radii, is very important since there are no valence
strange quarks in the nucleon. 
The aforementioned contributions to the properties of the nucleon are all 
included in its strange form factors, which means that 
the question may be rephrased as to whether or not the strange form factors
contribute to e.g. the nucleon electromagnetic form factors.
Since it became clear that such question could be addressed in 
weak neutral current experiments~\cite{KM88,M94}
a great deal of effort both theoretical and experimental has been put into this aspect 
of the strange content of the nucleon.
%
% Paragrafo

On the experimental side examples of the ongoing  
efforts focusing on the strange vector form factors are the measurements
by the SAMPLE~\cite{SAMPLE97,SAMPLE00s} and HAPPEX~\cite{HAPPEX00} collaborations.
The most recent measurement by the SAMPLE collaboration~\cite{SAMPLE00s} for the 
strange magnetic form factor finds ($Q^2$ in (GeV/c)$^2$)
\begin{equation} 
        G_M^{\rm s} (Q^2=0.1) = ( + 0.14 \pm 0.29 \, ({\rm stat.}) \pm 0.31 \, ({\rm syst.})) 
\;\; \mbox{n.m.}\, .  
\label{eq:samplenew}
\end{equation}
It is obtained from knowledge both on the neutral weak magnetic 
form factor $G_M^Z$ measured in
parity-violating elastic electron-proton scattering and on the electromagnetic 
form factors $G_{M}^{p\gamma}$,  $G_{M}^{n\gamma}$ using the relation
\begin{equation}
       G_M^Z =  \left(1-4\sin^2\theta_W\right)  G_M^{p\gamma} 
                 - G_M^{n\gamma} - G_M^{\rm s}\, ,       
\label{eq:GMZ}
\end{equation} 
where $\theta_W$ is the weak mixing angle determined  
experimentally~\cite{PDG00}: $\sin^2\theta_W=0.23147$.
The HAPPEX collaboration also announced results on the measurement of the strange vector form 
factors~\cite{HAPPEX00}: From the parity-violating polarized electron scattering 
asymmetry $A_{\rm th}$ they arrive after various approximations at  
\begin{equation}
                \frac{(G_E^{0} + 0.392 G_M^{0})}{(G_M^{p\gamma }/\mu_p)} (Q^2=0.477)
        =  1.527\pm 0.048\pm 0.027\pm 0.011.
\label{eq:happex0}
\end{equation}
The first error is statistical, the second is systematic and the last
comes from the uncertainty of the proton axial form factor~\cite{HAPPEX00}.
The available data for the electromagnetic form factors have access
to the strange vector form factors through 
\begin{equation}
                G_{E,M}^{\rm s}  =  G_{E,M}^0  -  G_{E,M}^{p\gamma}  -  G_{E,M}^{n\gamma}
\label{eq:GEMs}
\end{equation}
with the result
\begin{equation}
                 (G_E^{\rm s} + 0.392 G_M^{\rm s})(Q^2=0.477)  =  0.025\pm 0.020\pm 0.014.
\label{eq:happexs}
\end{equation}
Here, the first error is due to the errors in $G_0$ combined in
quadrature and the second one arises from the electromagnetic form factors.
Both procedures (\ref{eq:GMZ},\ref{eq:GEMs}) show how uncertainties in the knowledge 
of the electromagnetic form factors propagate into the error bars of the results 
(\ref{eq:samplenew},\ref{eq:happexs}) for the strange form factors. 
In particular, the HAPPEX results show a much smaller error bar
at the level of (\ref{eq:happex0}), which is extracted directly from the asymmetry 
$A_{\rm th}$, than (\ref{eq:happexs}), obtained from Eq.~(\ref{eq:happex0})
using Eq.~(\ref{eq:GEMs}). 
Unfortunately in the chiral quark-soliton model ($\chi$QSM) we are presently not yet able to 
calculate the axial-vector form factor to the same technical level as the 
electromagnetic ones.
Thus we cannot calculate $A_{\rm th}$ but will focus on the 
expressions~(\ref{eq:happex0}) and (\ref{eq:happexs}) for HAPPEX. 
%
%Paragrafo

On the theoretical side there have been many attempts to predict 
the strange form factors, differing by the    
methods used to study the non-perturbative physics behind them:
dispersion relations~\cite{J89,F97,HR98},
Skyrme model~\cite{PW92},
vector dominance models with $\omega-\phi$ mixing~\cite{FNJC94},
Nambu--Jona-Lasinio soliton model~\cite{WAAR95},
chiral bag model~\cite{HPM97},
meson exchange model~\cite{MMSO97},
heavy baryon chiral perturbation theory~\cite{HMS98,HKM99} and a
chiral quark model~\cite{HRG00}.
The different physics going into these approches explains
the wide range of predictions obtained.
Only the experimental data expected in the near future 
will allow to clarify the role of the mechanisms underlying these approaches.
The non-perturbative lattice QCD calculations point towards a negative 
strange magnetic moment~\cite{M00}, but are not yet conclusive regarding 
the exclusion of positive values~\cite{LT00}.
The strange form factors have already been investigated within 
the framework of the $\chi$QSM~\cite{KWG97}.
However, it was later found that the collective SU(3) quantization 
used there for the $\chi$QSM did not fulfil the Gell-Mann--Nishijima relation exactly,
a fact which is caused by the inherent 
nonlocality in time of the collective operators in 
the $\chi$QSM and the asymmetric transitions between the solitonic states 
and vacuum ones.
We use in this study a procedure known as the symmetry-conserving 
quantization~\cite{PWG98} for which the above problem does not arise. 
This new quantization has implications mainly at the level of the
magnetic moments and hence is relevant for the present study.

\medskip
%222222222222222222222222222222222222222222222222222222222222

{\bf 2.}   The real Dirac and Pauli vector form factors are defined from
           the matrix elements of the quark current $J_{\mu }$.   
In standard notation  
\begin{equation}
       \langle N (p')|J_{\mu }| N(p)\rangle 
                \; = \; \bar{u}_{N}(p')
                      \left[ \gamma _{\mu }F_{1}(Q^{2})
                           + i\sigma _{\mu \nu }\frac{q^{\nu }}{2M_{N}}F_{2}(Q^{2})
                       \right] u_{N}(p),
\label{eq:ff1}
\end{equation}
where $M_{N}$ and $u_{N}(p)$ are the nucleon mass and the nucleon spinor, respectively, and
$Q^2=-q^2>0$ is the square of the space-like four-momentum transfer, $q=p'-p$. 
In the present work we are interested in the form factors for the strange 
and baryonic currents. The strange quark current 
\begin{equation}
                     J^{\rm s}_{\mu }\; = \; \bar{\rm s}\gamma _{\mu }s
                                 \; = \; J^{B}_{\mu }-J^{Y}_{\mu }
\label{eq:scur}
\end{equation}
is expressed in terms of the baryonic $J^{B}_{\mu }$ and the hypercharge $J^{Y}_{\mu }$
currents:
\begin{eqnarray}
          J^{B}_{\mu } & = & \frac{1}{N_{c}}\bar{q}\gamma _{\mu }q\, ,   
\label{eq:baryoniccurrent}              \\
          J^{Y}_{\mu } & = & \frac{1}{\sqrt{3}}\bar{q}\gamma _{\mu }\lambda _{8}q.
\label{eq:bar}
\end{eqnarray}
We prefer, however, to work with the electric and magnetic Sachs form factors, 
related to the previous ones by
$G_E = F_1-\tau F_2$, with $\tau=Q^{2}/(4M^{2}_{N})$, and $G_M = F_1 + F_2$.
Such a choice is more natural when working in the Breit frame and in the 
non--relativistic limit, as is our case, because the form factors are then expressed
in a simple way by the time and space components of the current
\begin{eqnarray}
    \langle N'(J'_3)p'|J^0(0)|N(J_3)p\rangle & = &  G_E(Q^2)\delta_{J'_3,J_3},\label{eq:ge}  \\
    \langle N'(J'_3)p'|J^k(0)|N(J_3)p\rangle 
                        & = &  \frac{i}{2M_N}\epsilon^{klm}(\tau^l)_{J'_3J_3}q^m G_M(Q^2).
\label{eq:gm} 
\end{eqnarray}
The matrix element $\langle N'(J'_3)p'|J^\mu(0)|N(J_3)p\rangle$ is
conserved in the $\chi$QSM for the calculation of form factors
described below.
The form factors $G_{E,M}^0$ and $G_{E,M}^{\rm s}$ here presented refer to the currents
$J^B_\mu$ (\ref{eq:baryoniccurrent}) and $J^{\rm s}_\mu$ (\ref{eq:scur}), respectively. 

\medskip
%333333333333333333333333333333333333333333333333333333333333

{\bf 3.} The starting point for the $\chi$QSM~\cite{DPP88}  
         (for more details see~\cite{Ch96,ARW96})
         is the low-energy partition function in Euclidean space given by the functional 
integral over pseudo-Goldstone meson ($\pi^a$) and quark fields ($\psi$)
\begin{equation}
{\mathcal{Z}}  =  \int\! {\mathcal{D}} \psi {\mathcal{D}}\psi^{\dagger }{\mathcal{D}}\pi
                  \:\exp \left[-\!\int\!d^{4}x\psi^\dagger_f D(\pi)_{fg}\psi_g\right]
               =  \int\! {\mathcal{D}}\pi\:\exp (-S_{\rm eff}[\pi ]),
\label{eq:action}
\end{equation}
where $S_{\rm eff}[\pi]$ denotes the effective action 
\begin{equation}
                 S_{\rm eff}[\pi ] \; =  \; -N_c {\rm Tr}\, {\rm ln}\, D(\pi)\, .
\end{equation}
In these expressions ${\rm Tr}$ stands for the functional trace, 
$N_c$ the number of colors, and   
$D$ the Dirac differential operator in Euclidean space
\begin{equation}
      D  \; =  \; \gamma_4\left( \rlap {/}{\partial }-\hat{m}-MU^{\gamma5 } \right)
         \; =  \; \partial_4 + h(U) + m_8\gamma_4\lambda^8 \, .
\label{eq:diracd}
\end{equation}
The $\lambda^a$ are the Gell-Mann matrices normalized as 
${\rm tr}(\lambda^a\lambda^b)=2\delta^{ab}$ and $\hat{m}$ is the
matrix of the current quark mass given by $\hat{m}= \mbox{diag}(\bar{m},\bar{m},m_{\rm s})$
$=m_1\bm{1}+m_8\lambda^8$ with $m_1=(2\bar{m}+m_{\rm s})/3$ 
and $m_8=(\bar{m}-m_{\rm s})/\sqrt{3}$ 
assuming isospin symmetry, {\em i.e.} with $\bar{m}=m_{\rm u}=m_{\rm d}$. 
The constituent quark mass $M$ arises from the spontaneous breakdown of
chiral symmetry and is considered to be constant and the only free parameter of the model. 
The proper-time regularization is chosen in order to complete the definition of the model. 
For $U$ we assume a structure corresponding to Witten's so-called trivial embedding of 
SU(2) into SU(3): 
\begin{equation}
               U_{\rm SU(3)}\; =\; \left( \begin{array}{cc}
                             U_{\rm SU(2)} & 0        \\
                                          0 & 1         \end{array}\right) ,
\label{eq:imbed}
\end{equation}
with the SU(2) hedgehog given by
\begin{equation}
                  U_{\rm SU(2)}\; =\; \exp {[i\gamma_5 {\bf n}\cdot {\bm \tau }P(r)]}\, .
\label{eq:profile} 
\end{equation}
The partition function ${\mathcal{Z}}$ (\ref{eq:action}) can not be treated 
exactly as far as the integration over the pseudoscalar fields is concerned and thus requires  
some approximation framework, chosen in this model to be that of a large number of 
colors $N_c$.  
In the large $N_{c}$ limit the Euclidean partition function ${\mathcal{Z}}$
can be evaluated in the saddle-point approximation, which corresponds at the 
classical level to finding the profile function $P(r)$ in Eq.~(\ref{eq:profile}) 
which makes the action stationary.
This means in practice the numerical solution of the functional fixed point equation 
resulting from $\delta S_{\rm eff}/\delta P(r)=0$. 
Such procedure yields a classical field $U_c$ built from a set of single 
quark energies and corresponding states, $E_{n}$ and $\Psi _{n}$, pertaining to the 
one-particle eigenvalue problem for the Hamiltonian $h(U)$ in Eq.~(\ref{eq:diracd}).  
This mean-field object does not have the quantum numbers of the nucleon state, however.
To obtain the nucleon quantum numbers it is necessary to extract them from $U_c$ by 
the semiclassical quantization of the rotational and translational zero-modes: 
The zero modes are treated exactly within the path integral formalism by 
introducing collective coordinates (see~\cite{Ch96} for further details.)
Explicitly, the rotational and translational zero modes are taken into
account by means of the replacement
\begin{equation}
       U({\bm x},t)\; = \; R(t)U_c({\bm x}-{\bm Z}(t))R^{\dagger }(t),
\end{equation}
{\em i.e.} by rotating the classical saddle point solution $U_c({\bm x})$ 
through a unitary time-dependent SU(3) collective orientation matrix $R(t)$ and by considering 
all the collective displacements by ${\bm Z}$ of the center of mass of the soliton 
in coordinate space.
The quantization of these collective coordinates will yield the collective wave functions
with the appropriate quantum numbers.
%
%  Paragrafo

The functional integral (\ref{eq:action}) is thus replaced by a functional integral
over $R$.
As described in Ref.~\cite{Ch96}, we assume adiabatic rotation and small 
collective velocities in the 
calculation of the matrix elements for the form factors.
This means e.g. that in the expansion of
\begin{eqnarray}
\lefteqn{ D^{-1}\Bigl(R(t)U_c({\bm x}-{\bm Z}(t))R^{\dagger }(t)\Bigr) = }   \nonumber \\
          &=& e^{i{\bm Z}\cdot{\bm P}}R
              \left[D(U_c) +  R^{\dagger }\dot{R}
                     +\gamma _{4} R^{\dagger }m_8\lambda^8 R
              \right]^{-1} 
              R^{\dagger }e^{-i{\bm Z}\cdot{\bm P}}
\label{eq:prop}
\end{eqnarray}
only terms up to linear order
in the angular velocity $R^{\dagger }\dot{R}$ and $m_8$ are retained.

\medskip
%444444444444444444444444444444444444444444444444444444444444444444444

{\bf 4.}  A detailed account on how the electromagnetic form factors are derived 
          within this framework is given in Ref.~\cite{KBPG98}. 
Here we would like to point  out that the only free parameter of the model is the 
constituent quark mass $M$. 
We use in this study the value $420$~MeV,
which is found from a best fit to many nucleon observables~\cite{Ch96}. 
The other parameters of the model are the current nonstrange quark mass and 
the cut-off parameter of the proper-time regularization: 
They are fixed for a given $M$ in the mesonic sector by requiring the
physical pion mass and decay constant to be reproduced by the model. 
The mass of the strange quark is throughout this work set to $180$ MeV.
%
%  Paragrafo 

%
%   Fig. 1
%
\begin{figure}\begin{center}\includegraphics[height=6cm]{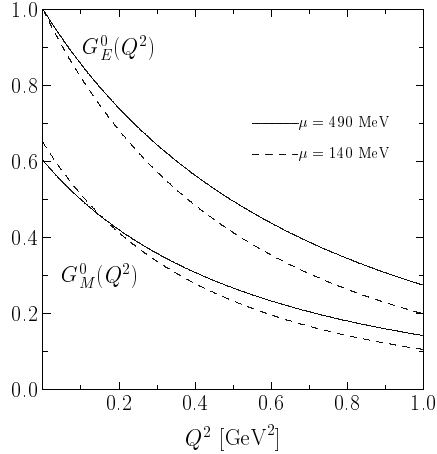}
\caption{The form factors $G_E^0$ and $G_M^0$ (in physical n.m.) as a function of $Q^2$.
         The solid curve and dashed curves represent the results for the kaon ($\mu=490$ MeV) 
         and pion ($\mu=140$ MeV) asymptotic tails, respectively. 
         The constituent quark mass is $M=420$ MeV. }
\end{center}\end{figure}
In Fig. 1 we present the results for $G_E^{0}$ and $G_M^{0}$. 
These figures show a general effect which is found in all our calculation, namely
the influence of the Yukawa mass $\mu$, governing the asymptotic behavior 
of the profile function at large $r$, $\exp(-\mu r)/r$, on the form factors.
While the SU(2) soliton incorporates the asymptotic pion behavior 
$\exp(-m_\pi r)/r$ naturally,
the construction of the SU(3) hedgehog by the embedding the SU(2)
soliton (\ref{eq:imbed}) forces all other pseudo-Goldstone bosons 
to share the same asymptotic behavior.  
Therefore, this treatment is phenomenologically unsatisfactory. 
It is expected that the 
kaon influence on the asymtotics of the meson fields play an important 
role in the description of the hyperons, even more so if they contain valence 
strange quarks, as well as in the specific contribution of the strange quark to
the properties of the nucleon.    
With this in mind we tentatively implement a heavier mass asymptotic behavior 
of the profile function $P(r)$ (\ref{eq:profile}) by adding a diagonal mass term 
to the Hamiltonian in Eq.~(\ref{eq:diracd}) and modifying the 
perturbative treatment of $m_s$ (or $m_8$) in Eq.~(\ref{eq:prop}) 
by subtracting the same term.
We prefer to look at the interval spanned by results of the more consistent pion 
asymptotics and the phenomenologically driven kaon asymptotics of $P(r)$ as giving an idea 
of the systematic model uncertainties steming from the lack of exact treatment of 
the SU(3) meson asymptotics.   
Indeed, it is found that the model nucleon electromagnetic form factors do not 
appreciably change under such procedure, except for the strange electromagnetic 
form factors and the electric form factor of the neutron.
%
%  Paragrafo

In Table~1 we present the results for the combination in Eq.~(\ref{eq:happex0}) 
of $G^0_E$ and $G^0_M$ and for the combination
of $G^{\rm s}_E$ and $G^{\rm s}_M$ in Eq.~(\ref{eq:happexs}) 
as given by HAPPEX~\cite{HAPPEX00}. 
Table~1 further distinguishes between  the cases
with the HAPPEX collaboration 
phenomenology based on value $\beta=0.392$
and the model calculation of the
quantity $\beta$, as given by
\begin{equation}
         \beta(Q^2,\theta)=\frac{\tau  G^{p\gamma}_M}{\epsilon  G^{p\gamma}_E}
\label{eq:beta}
\end{equation}
( with $\epsilon=[1+2(1+\tau)\tan^2(\theta)]^{-1}$ and $\tau=Q^{2}/(4M^{2}_{N})$),
using the model electromagnetic form factors
calculated with the symmetry-conserving quantization.
For the combination of $G^0_E$ and $G^0_M$ and 
for both cases of $\beta$ the HAPPEX result~(\ref{eq:happex0}) falls within 
the interval of model results between the pion ($\pi$) and kaon ($K$) asymptotics.  
The combination of $G^{\rm s}_E$ and $G^{\rm s}_M$ seems to overestimate the HAPPEX 
result~(\ref{eq:happexs}) for both cases of $\beta$. 
One notices that the kaon asymptotics leads to results closer to the experiment.
One has to take into account, however,
that the HAPPEX value~(\ref{eq:happexs}) could be higher, by as much as $0.020$,
with a different input for the electromagnetic form factors~\cite{HAPPEX00}.   
Although the results in Table 1 depend on $M$ it is found that,
for physically reasonable values of $M$ between $400$ and $450$ MeV
the dependence on $M$ is much weaker than the effect of the meson asymptotics. 
%
%
%      Tab. 1  
%
\begin{table}[t]\begin{center}\begin{tabular}{|c|c|c|c|c|}               \cline{2-5} 
\multicolumn{1}{c}{ } &
 \multicolumn{2}{|c|}{$(G^0_E+\beta(Q^2,\theta)G^0_M)/(G^p_M/\mu_p)$
 }&   \multicolumn{2}{c|}{$G^{\rm s}_E+\beta(Q^2,\theta)G^{\rm s}_M$ }     \\ \hline
 $\beta(Q^2,\theta)$  &      $0.392$                     &     $0.308$ (Model)               &      $0.392$                     &     $0.308$ (Model)     \\ \hline
$\mu$                 &      $\pi$   -  K                &     $\pi$   -  K                  &      $\pi$   -  K                &     $\pi$   -  K        \\ \hline
$M=420$ MeV           &      $1.433$ - $1.695$           &     $1.377$ - $1.621$             &      $0.103$ - $0.071$           &     $0.101$ - $0.067$   \\ \hline
HAPPEX                &    \multicolumn{2}{|c|}{$1.527\pm0.048\pm0.027\pm0.011$}             &     \multicolumn{2}{c|}{$0.025\pm0.020\pm0.014$ }          \\ \hline     
\end{tabular}\end{center}
\caption{The combinations $(G^0_E+$$\beta(Q^2,\theta)G^0_M)$$/(G^{p\gamma}_M/\mu_p)$ and 
         $G^{\rm s}_E+$$\beta(Q^2,\theta)G^{\rm s}_M$ for the HAPPEX kinematics $Q^2=0.477$ GeV$^2$ 
         and $\theta=12.3^\circ$.  The experimental data are taken from HAPPEX~\cite{HAPPEX00}.}
\end{table}
%
% Paragrafo

The results for the strange electric form factor $G_{E}^{\rm s}$ are shown in Fig.~2 and 
for the strange magnetic form factor $G_{M}^{\rm s}$ in Fig.~3.
%
% Fig. 2
%
\begin{figure}\begin{center}\includegraphics[height=6cm]{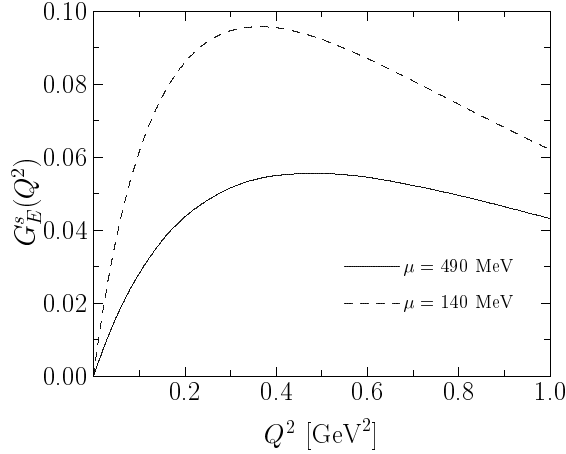}
\caption{The strange electric form factor $G_E^{\rm s}$ as a function of $Q^2$.
         Conventions and model parameter as in Fig.~1.}
\end{center}\end{figure}
%
% Fig. 3
%
\begin{figure}\begin{center}\includegraphics[height=6cm]{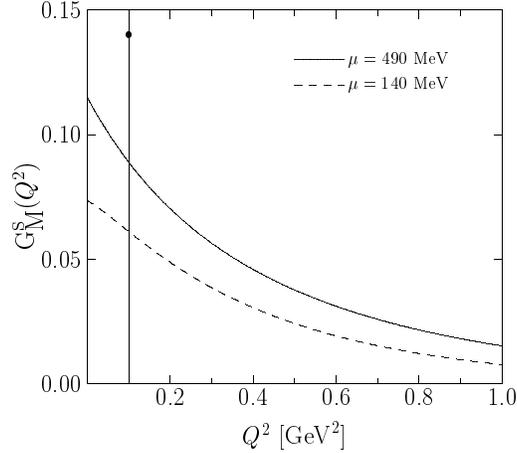}
\caption{The strange magnetic form factor $G_M^{\rm s}$,  in units of
         the physical n.m., as a function of $Q^2$. 
         Conventions and model parameter as in Fig.~1.  The
         experimental data are taken from SAMPLE~\cite{SAMPLE00s}}
\end{center}\end{figure}
For $G_E^{\rm s}$ the result is basically the same as in Ref.~\cite{KWG97}
while the $G_M^{\rm s}$ is strongly modified, compared to~\cite{KWG97}, to a
larger extent by removing a numerical error, as reported in
Ref.~\cite{KPPG98}, and to a lesser extent by using in the present
investigation the scheme of the symmetry-conserving quantization.
Only in the case of $G_M^{\rm s}$ is a direct comparison with experiment possible. 
The SAMPLE value (\ref{eq:samplenew})~\cite{SAMPLE00s}
is shown in Fig.~3: The model value is not far from the experimental one, but the error 
bar is too big to draw any other conclusion.
The fact that a positive value is obtained, and in particular that the strange magnetic
moment turns out positive, is in agreement with the model independent analysis 
of~\cite{KPPG98}, which finds $\mu_s=(0.41\pm0.18)$ n.m. on the basis of the SU(3)
model algebra with input from the hyperon magnetic moments.
%
%  Paragrafo

In Table~2 we list the $\chi$QSM results for the strange 
electric and magnetic radii, defined as
\begin{equation}
     \langle r^2\rangle^{\rm s}_E \; = \; -6\left.\frac{dG^{\rm s}_E(Q^2)}{dQ^2}\right|_{Q^2=0},
\end{equation}
\begin{equation}
     \langle r^2\rangle^{\rm s}_M \; = \; -\frac{6}{\mu_{\rm s}}
               \left.\frac{dG^{\rm s}_M(Q^2)}{dQ^2}\right|_{Q^2=0}  ,
\end{equation}
and the strange magnetic moment, for $M=420$ MeV and both the pion ($\pi$) and 
kaon ($K$) asymptotics. 
%
% Tab. 2
%
\begin{table}[t]\begin{center}\begin{tabular}{|l|rr|}                \hline
$\mu$                             &    $\pi$    &          K      \\ \hline
$\langle r^2\rangle^{\rm s}_E$ [Fm$^2$] &   $-0.220$  &     $-0.095$    \\ \hline
$\mu_s$ [n.m.]                    &   $ 0.074$  &     $ 0.115$    \\ \hline
$\langle r^2\rangle^{\rm s}_M$ [Fm$^2$] &   $ 0.303$  &     $ 0.631$    \\ \hline
\end{tabular}\end{center}
\caption{The strange magnetic moment and the mean-square strange radii for $M=420$ MeV.}
\end{table}
%
% Paragrafo

Using Eq.~\ref{eq:GMZ} it is possible to obtain the model prediction for $G^Z_M$,
which is given in Fig.~4, where the experimental result~\cite{SAMPLE00s}
\begin{equation}
     G_M^Z(Q^2=0.1) = ( 1.49\pm 0.29 \,({\rm stat.}) \pm 0.31\, ({\rm syst.}))\; 
{\rm n}.{\rm m}.\, .
\end{equation}
is also plotted. 
Again the $\chi$QSM result falls within the (admittedly large) error bars.
%
%
% Fig. 4
%
\begin{figure}\begin{center}\includegraphics[height=6cm]{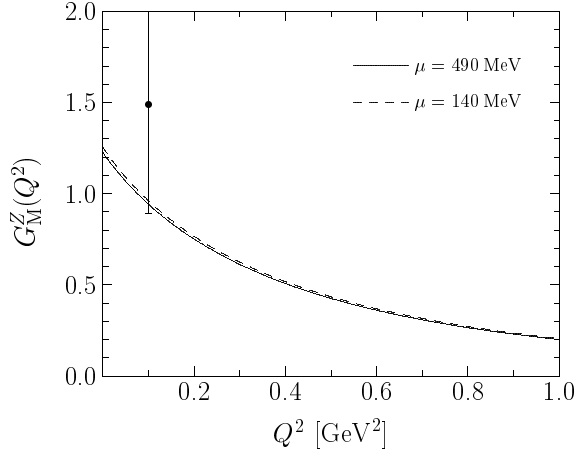}
\caption{The neutral magnetic form factor $G_M^{\rm s}$,  in units of
         the physical n.m., as a function of $Q^2$. 
         Conventions and model parameter as in Fig.~1.  The 
         experimental data are taken from SAMPLE~\cite{SAMPLE00s,Pitt}.}
\end{center}\end{figure}
%
%  Paragrafo

Lastly, Table~3 shows the full model predictions, {\em i.e.} using both the strange
form factors and the electromagnetic ones, for the kinematics of future 
HAPPEX and A4 experiments.    
This table is made for the the combination $(G_E^{\rm s} + $$\beta
G_M^{\rm s} )$
$/(G_M^{p\gamma}/\mu_p)$ of form factors because it is closer to 
the quantities extracted from
the asymmetries and it can reliably be calculated in the model since this describes well
the $Q^2$ dependence of the form factors~\cite{PWG98,Ch96}.  
For the sake of consistency we calculate the factor $\beta$ as given by
Eq.~(\ref{eq:beta}) using the model.  
Again we see that the pion asymptotics deviates more from the data 
than the kaon one, which, nevertheless, seems to overestimate the data.
Apart from having to rely on more data to confirm whether this trend will persist,
one has to bare in mind that the strange form factors as extracted with 
the help of Eq.~(\ref{eq:GMZ}) contain a piece due to isospin symmetry 
breaking~\cite{Ma97}. 
The isospin symmetry breaking effects were considered here to be small on
account of large $N_c$. 
%
% Tab. 3
%
\begin{table}[t]\begin{center}\begin{tabular}{|c|c|c|c|c|c|c|}                                                   \hline 
Exp.               &     A4             &  HAPPEX II         &  A4                &   HAPPEX            &  A4               \\ \hline
$Q^2$(GeV$^2$)     &    $0.10$          &  $0.11$            &  $0.23$            &   $0.48$            &  $0.48$           \\ \hline 
$\theta$($^\circ$) &    $35$            &  $6$               &  $35;145$          &   $12.3$            &  $145$            \\ \hline 
$\beta$ (Mod.)     &    $0.105$         &  $0.059$           &  $0.255$           &   $0.308$           &  $0.587$          \\ \hline 
$\mu$              &  $\pi$  -K         &  $\pi$  -  K       &  $\pi$  -  K       &   $\pi$  -  K       &  $\pi$  -  K      \\ \hline
$M=420$ MeV        &  $0.088$-$0.048$   &  $0.091$-$0.046$   &  $0.174$-$0.108$   &   $0.278$-$0.185$   &  $0.297$-$0.213$  \\ \hline
\end{tabular}\end{center}
\caption{The values of $(G^{\rm s}_E+$$\beta(Q^2,\theta)G^{\rm s}_M)$$/(G^{p\gamma}_M/\mu_p)$ 
         at various $Q^2$ and $\theta$ for $M=420$ MeV.
         with pion ($\pi$) and kaon (K) tails. 
         All the experiments (Exp.) are still being performed except for HAPPEX.} 
\end{table}

\medskip
%555555555555555555555555555555555555555555555555555555555555555

{\bf 5.} In this work, we have investigated the strange properties of
         the nucleon: we calculated $G_{E,M}^0$, the strange form
         factors $G_{E,M}^{\rm s}$
and their associated radii, and the strange magnetic moment $\mu_s$ within the framework 
of the SU(3) $\chi$QSM.  
On the whole, the results of the model calculation with kaon asymptotics turn 
out to be in better agreement with experiment than those with the pion asymptotics,
as expected for strange quantities.  
The results for the kaon asymptotic turn out to be in fairly good agreement with 
the available data, apart from an  apparently
slight overestimation of the available HAPPEX data till now on
$G^{\rm s}_E+$$\beta G^{\rm s}_M$.
The difference between kaon and pion asymptotics gives an indication 
of the systematic error of the $\chi$QSM in calculating 
strange quantities.
Future data from SAMPLE, HAPPEX, and also from the G0 (JLAB) and A4 (MAMI, Mainz) 
collaborations, will allow one to judge further the results of the chiral
quark-soliton model.

\bigskip
%%%%%%%%%%%%%%%%%%%%%%%%%%%%%%%%%%%%%%%%%%%%%%%%%%     acknowledgements
\noindent{\large\bf Acknowledgments}

\medskip

\noindent 
The authors are grateful  to thank F. Maas (MAMI A4) and M. Pitt
      (SAMPLE) for clarifying experimental 
      related points and to P. Pobylitsa, M. Polyakov, M. Prasza\l owicz, and
      M. Vanderhaeghen for useful discussions and critical comments.  
AS acknowledges partial financial support from Praxis XXI\-/BD\-/15681\-/98.
The work of HCK is supported by the Korean Research Foundation (KRF\--2000\--015\--DP0069). 
This work has partly been supported by the BMBF, the DFG,
       the COSY--Project (J\" ulich), and POCTI\-/32845\-/FIS\-/2000.

%     bibliography

\end{document}